awk 'BEGIN{skip=100000}/.*
     {skip--;if(skip==-1)print "Writing",a[2];if(skip<0)print $0 > a[2]}'

\documentstyle[preprint,revtex]{aps}
\begin{document}

\begin{title}
Elastic String in a Random Potential
\end{title}

\author{M. Dong, M.C. Marchetti, A. Alan Middleton}
\begin{instit}
Physics Department, Syracuse University, Syracuse, NY 13244
\end{instit}

\author{V. Vinokur\cite{AAAuth}}
\begin{instit}
Argonne National Laboratory, Materials Science Division, Argonne, IL 60439
\end{instit}
\receipt{DRAFT \today}

\begin{abstract}
We have studied numerically the dynamics of a directed
elastic string in
a two-dimensional array of quenched random impurities. The string is
driven by a constant transverse force and thermal fluctuations are
neglected. There is a transition from pinned to unpinned behavior at
a critical value $F_T$ of the driving force.
At the transition the average string velocity
scales with the driving force. The scaling is equally well described by
a power law
$v_d\sim (F-F_T)^\zeta$, with $\zeta=0.24\pm0.1$, or by
a logarithm, $v_d\sim1/\ln(F-F_T)$. The divergence
of the velocity-velocity correlation length at threshold
is characterized by an
exponent $\nu=1.05\pm0.1$.
\end{abstract}

The effect of disorder on the dynamical properties of an overdamped
elastic string driven by a constant force is of interest for a number
of physical situations. These include one-dimensional models of charge
density waves (CDW) \cite{fukuyama,sneddon}, flux flow
in type-II superconductors \cite{larkin}
and the motion or growth of various linear boundaries
between two systems or two phases of the same system,
such as the interface of a fluid displacing a second fluid in a
porous medium \cite{rubio}. This model is also closely related to
models of friction, earthquakes, and sandpiles \cite{soc}.

The specific problem considered here,
the motion of an elastic string driven by a transverse constant force
through a two-dimensional array of randomly-distributed pinning centers,
has been studied analytically in various limits
\cite{larkin,feigelman,fisher,ioffe}.
Feigel'man \cite{feigelman} used
perturbation theory in the disorder
to analyze the motion at large driving forces, at both zero
and finite temperature. Ioffe and Vinokur discussed the
string dynamics for small driving forces and finite
temperature \cite{ioffe}.
While the string dynamics in these asymptotic
regimes is rather well understood,
the dynamics near threshold
needs further investigation.

In this paper we present the results of a numerical study
of the motion of the string
in the absence of thermal fluctuations.
We have found that, as in the CDW model, there is a threshold value of
the driving force below which the string is stuck in a pinned configuration
and above which the string moves on the average in the direction
of the driving force.
As discussed below, there are, however, qualitative and quantitative
differences between CDW models and the model we study here.
We are mainly interested in the behavior just above
this depinning transition, where transport is
collective and perturbative methods cannot be used.
The behavior near threshold
is that of a dynamic critical phenomenon and it can be
described by scaling laws and critical exponents \cite{fisher}.
We find that the average velocity scales with the reduced driving force
$f=(F-F_T)/F_T$.
The data is equally well described by a
logarithm, $v_d\sim1/\ln(f)$, or by a power law,
$v_d\sim f^{\zeta}$, with a small exponent $\zeta=0.24\pm0.1$.
To gain insight into the behavior near threshold,
we have also studied
equal-time velocity and position correlation functions.
The spatial range of the velocity correlations
is determined by a correlation length $\xi$ that
diverges near threshold as $\xi\sim f^{-\nu}$, with $\nu=1.05\pm0.1$.
We find that the temporal fluctuations around the mean velocity
also diverge as threshold is approached, with an exponent consistent
with that found from the scaling of the velocity correlations
and arguments due to Fisher \cite{fisher}.

The main difference between CDW simulations in one dimension
and the model considered here is in the pinning potential.
In CDW's the pinning potential is uncorrelated along
the length of the string, but
is periodic in the direction transverse to the string.
Here the pinning potential has short-range
correlations both along the string direction and transverse to the
string.
This model is relevant for the motion
of magnetic flux lines
in the mixed phase of type-II superconductors.
Because the disorder is uncorrelated, any finite string will eventually
encounter a region of exceptionally strong pinning that will stop
its motion.
However,
the expected time to encounter such a region increases exponentially
with system size.

The specific model we have studied is an elastic string embedded
in two dimensions and
on the
average aligned with the $z$ direction (in the superconductor this is the
direction of the applied magnetic field). The position of a point on the
string at height $z$ and time $t$ is denoted by $x(z,t)$.
The Hamiltonian for the string is
\FL
\begin{equation}
\label{eq:hamilt}
H=\int_0^Ldz \bigg\{
{K\over 2}\Big({\partial x\over\partial z}\Big)^2
+\sum_{i=1}^{N_p}U_i({\bf r}-{\bf R}_i)
-{\cal F}x(z)\bigg\},
\end{equation}
where $L$ is the size of the system in the $z$
direction.
The first term in Eq.\ (\ref{eq:hamilt}) is the elastic
energy of the string,
with $K$ the string elastic constant.
The second term describes the interaction with $N_p$ impurities
randomly distributed throughout the plane at positions ${\bf R}_i$.
Here ${\bf r}=(x(z),z)$ denotes a point on the string.
The interaction $U_i$ of the string with
the $i$-th pin is approximated by a potential
well centered at the pin location ${\bf R}_i$ with maximum depth $U_0$
(the depth is the
same for all the pins - it is not a random variable) and width
$R_p$ in both the $x$ and the $z$ directions.
The last term in Eq.\ (\ref{eq:hamilt})
arises from the constant driving
force per unit length ${\cal F}$ applied to the string in the $x$ direction.

We consider the overdamped string dynamics.
Neglecting thermal fluctuations, the equation of motion for the
elastic string is
\begin{equation}
\label{eq:rmotion}
{\partial x\over\partial t}={\partial^2 x\over\partial z^2}
-F_p\sum_{i=1}^{N_p}{\delta(U_i/U_0)\over\delta x}+F,
\end{equation}
where we have introduced
dimensionless variables and parameters.
All lengths are measured in units of $R_p$ and time is measured in units
of $t_0=\gamma R_p^2/K$, with $\gamma$ a friction coefficient.
For magnetic flux lines the friction
is the damping arising from interaction of the
normal core electrons with the crystal lattice of the superconductor
\cite{kim}.
Also $F={\cal F}R_p/K$ is a dimensionless driving force and $F_p=U_0/K$
a dimensionless pinning force.

The overall motion of the string can be described in terms of
a ``center-of mass" velocity,
defined as
\begin{equation}
\label{eq:cmvel}
v_{cm}({\cal F},t)=
    {1\over L}\int_0^Ldz~v\big(z,t\big),
\end{equation}
where $v\big(z,t\big)=\partial x/\partial t$ is
the instantaneous velocity
of a point on the string.
The center-of-mass velocity is a fluctuating quantity since
it depends on the
random positions of the pinning centers.
The drift or average velocity
of the string is given by
\begin{equation}
\label{eq:veloc}
v_d({\cal F})=<v_{cm}({\cal F},t)>,
\end{equation}
where the angular brackets denote the average over
time. In the calculation we average over impurity realizations
by performing an average over time
since as time evolves the string
samples different impurities configurations \cite{footnotea}.

We integrated numerically the discretized
version of the equation of motion (\ref{eq:rmotion}) for
strings composed of discrete elements, each
of dimensionless size $1$ (i.e., $R_{p}$)
in the $z$-direction, that interact through
nearest neighbor elastic forces.
We imposed periodic boundary conditions in the $z$-direction.
Each element is subject to an $x$- and $z$-dependent
pinning potential consisting of randomly
distributed triangular wells of width 1.
We have run simulations for two values of the well depth $U_{0}=0.05$
($F_p=0.1$) and $U_{0}=0.5$ ($F_p=1$);
the density of pins was chosen be $\rho=n_pR_p^2=0.1$, while the
elastic constant was unity. The first set of parameters
($F_p=0.1$) lies at the crossover between weak and
strong pinning regimes according to a dimensional estimate
of the condition for weak pinning, $\rho>F_p$.
By examining string configurations as a function
of time it appears, however, that this choice of parameters
corresponds to weak pinning.
The second set of parameters
($F_p=1$) lies in the strong pinning regime.
As the pinning force is discontinuous, we used a simple Euler
algorithm for the simulation
and chose a time step small enough (typically $\Delta t = 0.1$)
that the results were insensitive to doubling the time step.
For fields within 10\% of threshold, the length of the simulation usually
exceeded $4 \times 10^7$ time steps, while
for fields further from threshold, shorter simulation times
gave very good averages.
We investigated systems of sizes from 256 to 16384.

For $F_p=1$ the transition from pinned
to unpinned regime occurs at a critical value
$F_T=0.3058\pm0.0002$
of the driving force, as shown in Fig. 1.
This value is of the order given by a dimensional estimate,
$F_T=\rho F_p$, which gives $F_T=0.1$.
For $F_p=0.1$ the depinning transition takes place at
$F_T\simeq 0.01505$. This value is consistent with
the dimensional estimate of $F_T$ for both
strong ($F_T=\rho F_p$) and weak
($F_T=\rho^{2/3} F_p^{4/3}$) pinning,
since they both give $F_T=0.01$.
For large driving forces, $F\gg F_T$, the effects of pinning is
negligible and
the string advances uniformly, with $v_d\approx F$.
The deviations
from this asymptotic form were studied
by a perturbation
expansion in $F/F_p$ \cite{sneddon,feigelman}, with the result
$v_d/F=1-CF^{-3/2}$, in agreement with our simulations \cite{footnoteb}.
As the driving force $F$ is lowered towards the threshold value
the motion of the string becomes more and more jerky. Near threshold
the perturbation theory breaks down.
The scaling of the average velocity with reduced driving
force is well described by $v_d\sim1/\ln(f)$ or by
a power law,
$v_d\sim f^{\zeta}$, with a small exponent
($\zeta=0.24\pm0.1$ for $F_p=1$ and $\zeta=0.34\pm 0.1$
for $F_p=0.1$).
The strong pinning data are shown in Fig. 1.
One may expect that the scaling exponents near threshold
should be the same for weak and strong pinning
since they describe behavior of the string at
large length scales. At present our results do not answer
this question conclusively.
The scaling exponents
obtained for the two values of the pinning force
are consistent within experimental error, but more work is
needed to properly address this point.

In both weak and strong pinning regimes there is a correlation length
$\xi$ that
diverges at threshold as $\xi\sim f^{-\nu}$, with $\nu$
the correlation length exponent.
Away from threshold
$\xi$ has the bare value $\xi_0$ and is the smallest length scale
over which the shape of the string varies,
$\xi_0\sim l_p=(F_p\sqrt{\rho})^{-2/3}$ for weak pinning
and $\xi_0\sim 1/\rho$ for
strong pinning.
Near threshold the string can get stuck
for some time in a region where pinning forces roughly balance the
driving force: a section of the string
is coherently pinned and its velocity vanishes. After a while the
string moves forward and jumps to a new configuration.
This behavior is displayed in Fig. 2, showing maps of
the local velocity of the string.
White regions are regions of zero velocity, while dark regions
are regions where the velocity is largest. The stationary pinned
regions become very large near threshold.
The correlation length $\xi$ is related to the linear
dimension of these regions through the exponents
$\eta$ and $\kappa$ discussed below.

When the string is in a sliding state (i.e., one with $v_d\not= 0$)
the ``center-of-mass" velocity
$v_{cm}(t)$ exhibits fluctuations
in time that become large near threshold.
To analyze the velocity fluctuations we have considered
the rms velocity,
$v_{rms}=\sqrt{<[v_{cm}(t)]^2>-<v_{cm}(t)>^2}$.
After initial transients, the probability of observing a
given velocity at any time is fit well by a Gaussian shape, with
mean $<v_{cm}>$ and standard deviation $v_{rms}$ (note that in
CDW models, distinct samples will have differing $v_{rms}$).
This supports our expectation that the time to stopping due
to rare regions grows exponentially with the system size.
Just above the transition from pinned to unpinned behavior,
where the velocity is small,
the velocity fluctuations
are of the same size as the velocity itself.
Naively the magnitude of the rms noise relative to the average
velocity is expected to scale as the number of correlation
lengths in the sample,
$v_{rms}/v_d \sim (\xi/L)^{1/2}$.
On the other hand velocity correlations are not perfectly
coherent within a length $\xi$ \cite{fisher}.
As a result the effective coherence length that determines the
rms noise may be smaller than $\xi$ and
$v_{rms}/v_d \sim (\xi^{4-\eta}/L)^{1/2}$.
Here $4-\eta$ can be thought of as an effective dimensionality,
with $4-\eta\leq d$ and $d$ the dimensionality (here $d=1$).
Using $\xi\sim f^{-\nu}$, our results for the ratio $v_{rms}/v_d$
are well described over more than
two decades in reduced field and a factor of 16
in system size by this form with
$\nu(4-\eta)/2=0.5\pm0.05$
for both $F_p=0.1$ and $F_p=1$.

Above threshold the correlation length $\xi$
characterizes the decay of velocity fluctuations.
The equal-time velocity correlation function is defined as
\begin{equation}
\label{eq:eqvel}
C_v(z)= <\overline{v(z,t)v(0,t)}>-<[\overline{v}(t)]^2>,
\end{equation}
where the overbar denotes the spatial average
over the length of the string.
We find that velocity correlations scale with $\xi$
according to
$C_v(z)/v_d^2\sim \xi^{4-\eta-1} G_v(z/\xi)$ \cite{fisher},
with $\nu=1.05\pm0.1$ and $\eta=3.12\pm0.1$, consistent with the
results obtained for $\nu$ and $\eta$ from the rms noise, as
required by
$(1/L)\int_0^L~dz C_v(z)=v_{rms}^2$.
The scaling of velocity correlations is shown in
Fig. 3 for $F_p=1$. The figure also shows that
the decay of correlations within a length $\xi$ is fit by
a power law over more than two decades, i.e.,
$G_v\sim z^{-\kappa}$, with $\kappa=0.5\pm0.05$.
A power-law decay of velocity correlations was
also reported by Sibani and Littlewood \cite{sibani}
for CDW's.
Much longer simulation lengths are required for the convergence
of the data at the smaller value of the pinning force ($F_p=0.1$).
At present the evidence for power law decay of velocity correlations
in this case is not as good as for the data displayed in Fig. 3.

To analyze the distortion of the string configuration
under the
competing action of disorder and elasticity, we have evaluated
the correlation function of the transverse positions of points
on the string at a distance $z$,
\begin{equation}
\label{eq:sqdis}
C_x(z)=
 <\overline{[x(z)-x(0)]^2}>.
\end{equation}
The size $l_{\perp}(z)$ of the transverse
fluctuations
of a length $z$ of string
is $l_{\perp}(z)=\sqrt{C_x(z)}$.
Near threshold, where the correlation length $\xi$ is of
the order of the system size,
we find that the transverse
correlation length $l_{\perp}(z)$ scales as
$l_{\perp}(z)\sim z^{\chi}$, with $\chi=0.97\pm 0.05$ for
both pinning strengths.
Since $\chi$ is near one, linear elasticity theory is
marginally self-consistent, in contrast with one-dimensional CDW
models \cite{coppersmith}; the strain will become large only
in exponentially rare regions.
Away from threshold $\xi$ is smaller than the
system size and there is evidence for a crossover from a regime
where $l_{\perp}(z)\sim z^{\chi}$,
for $z<\xi$
to a regime
where $l_{\perp}(z)\sim z^{1/2}$,
for $z\gg \xi$.
Experimental measurements of the roughening exponent for the interface
that forms when a fluid displaces another in a porous medium
give $\chi\simeq 0.73$ \cite{rubio}.
Values of $\chi$ smaller than 1 ($\chi\sim 0.5-1.$) have also
been obtained for stochastic growth
models for the propagation of an interface \cite{medina}.
The stochastic equation of motion used to model the growth of an interface
contains, however, a term proportional to $(\partial x/\partial z)^2$ which is
neglected in our model.

\bigskip
This work was conducted using the computational resources of
the Northeast Parallel Architectures Center (NPAC) at Syracuse
University. It was
supported at Syracuse by the National Science Foundation through
grants DMR87-17337 and DMR91-12330 and at Argonne
by the U.S. Department of
Energy, BES-Material Sciences, under Contract No. W-31-109-ENG-38,
and The Science and Technology Center for Superconductivity.

\figure{\label{fig1}
The scaling of the drift velocity $v_d\sim 1/\ln(f)$
for $F_p=1$ and various system
sizes. The inset shows $v_d$ versus $F$ to highlight the depinning
transition.}

\figure{\label{fig2}
Maps of string velocity for $f=0.01$ (top) and $f=0.076$ (bottom)
and $F_p=0.1$.
The vertical axis is time, while the horizontal axis is the position
$z$ on the string. Dark regions indicate where the velocity exceeds
0.01. Maps are shown for strings of size $L=4096$
evolving over a time interval $\Delta t =30$.}

\figure{\label{fig3}
The velocity correlation functions $C_v(z)/v_d^2$
for $F_p=1$ and six different values of $f$ collapse on one curve when scaled
as described in the text. The straight line shows the power law decay
of the correlations at short distances, $C_v\sim z^{-\kappa}$, with
$\kappa=0.5$.}


\begin{references}
\bibitem{footnotea}
We verified explicitly that at long times the coarse-grained velocity
of the string (coarse-grained over a length larger than the spacing
between impurities) is independent of $z$
by calculating the time average
of the velocity of a single point on the string.
\bibitem{footnoteb}
This correction is strictly correct only at intermediate fields.
At very high fields, there is a crossover to the single-particle
behavior $1-v/F \sim F^{-2}$ that depends on the range of the
potential correlations in the $z$-direction. We see the latter behavior
at fields 100 times larger than the depinning field.
\end{references}
\end{document}